\documentclass[twocolumn,nofootinbib,aps,physrev,groupedaddress,superscriptaddress,amsmath,amssymb,noeprint]{revtex4-2}
\usepackage[T1]{fontenc}
\usepackage{graphicx}
\graphicspath{{./Figures/}}
\usepackage{dcolumn}
\usepackage{bm}
\usepackage[dvipsnames]{xcolor}
\usepackage{hyperref}
\hypersetup{colorlinks,linkcolor={MidnightBlue},citecolor={MidnightBlue},urlcolor={MidnightBlue}}  

\usepackage{float}

\usepackage{physics}
\usepackage{amsfonts}
\usepackage{verbatim}
\usepackage{amsmath}
\usepackage{csquotes}
\usepackage{color}
\usepackage{soul}
\usepackage{amsthm}
\usepackage{bm}
\usepackage{graphicx}
\usepackage[normalem]{ulem}

\usepackage{verbatim}

\newenvironment{equations}
{\begin{equation}\begin{aligned}}
		{\end{aligned}\end{equation}\ignorespacesafterend}

\newcommand{\prt}[1]{\left(#1\right)}

\newcommand{\intmp}{\int_{-\infty}^{+\infty}}

\newcommand{\Schr}{Schr\"{o}dinger\ }

\newcommand{\dg}{\dagger}

\newcommand{\prtq}[1]{\left[#1\right]}

\newcommand{\prtg}[1]{\left\{#1\right\}}

\newcommand{\prtgB}[1]{\Bigg\{#1\Bigg\}}

\newcommand{\mcT}{\mathcal{T}}

\newcommand{\tpsi}{\tilde{\psi}}

\newcommand{\Id}{\mathbb{I}}

\newcommand{\hx}{\hat{x}}
\newcommand{\hp}{\hat{p}}
\newcommand{\hq}{\hat{q}}

\newcommand{\hn}{\hat{n}}

\begin{document}

\title{Exploring the Accuracy of Interferometric Quantum Measurements under Conservation Laws}

\author{Nicol\`{o} Piccione}
\email{nicolo'.piccione@units.it}
\affiliation{MajuLab, CNRS–UCA-SU-NUS-NTU International Joint Research Laboratory}
\affiliation{Centre for Quantum Technologies, National University of Singapore, 117543 Singapore, Singapore}
\affiliation{Department of Physics, University of Trieste, Strada Costiera 11, 34151 Trieste, Italy}
\affiliation{Istituto Nazionale di Fisica Nucleare, Trieste Section, Via Valerio 2, 34127 Trieste, Italy}

\author{Maria Maffei}
\affiliation{Dipartimento di Fisica, Università di Bari, I-70126 Bari, Italy}
\affiliation{INFN, Sezione di Bari, I-70125 Bari, Italy}

\author{Andrew N. Jordan}
\affiliation{The Kennedy Chair in Physics, Chapman University, Orange, CA 92866, USA}
\affiliation{Institute for Quantum Studies, Chapman University, Orange, California 92866, USA}
\affiliation{Department of Physics and Astronomy, University of Rochester, Rochester, New York 14627, USA}

\author{Kater W. Murch}
\affiliation{Department of Physics, Washington University, St. Louis, Missouri 63130, USA}

\author{Alexia Auff\`{e}ves}
\affiliation{MajuLab, CNRS–UCA-SU-NUS-NTU International Joint Research Laboratory}
\affiliation{Centre for Quantum Technologies, National University of Singapore, 117543 Singapore, Singapore}

\begin{abstract}
	
	A (target) quantum system is often measured through observations performed on a second (meter) system to which the target is coupled.
	In the presence of global conservation laws holding on the joint meter-target system, the Wigner-Araki-Yanase theorem and its generalizations predict a lower-bound on the measurement’s error (Ozawa’s bound).
	While practically negligible for macroscopic meters, it becomes relevant for microscopic ones.
	Here, we propose a simple interferometric setup, arguably within reach of present technology, in which a flying particle (a microscopic quantum meter) is used to measure a qubit by interacting with it in one arm of the interferometer. In this scenario, the globally conserved quantity is the total energy of particle and qubit.
	We show how the measurement error, $\varepsilon$, is linked to the non-stationary nature of the measured observable and the finite duration of the target-meter interaction while Ozawa's bound, $\varepsilon_{\mathrm B}$, only depends on the momentum uncertainty of the meter’s wavepacket.
	When considering short wavepackets with respect to the evolution time of the qubit, we show that $\varepsilon/\varepsilon_{\mathrm B}$ is strictly tied to the position-momentum uncertainty of the meter’s wavepacket and $\varepsilon/\varepsilon_{\mathrm B} \rightarrow 1$ only when employing Gaussian wavepackets.
	On the contrary, long wavepackets of any shape lead to $\varepsilon/\varepsilon_{\mathrm B} \rightarrow \sqrt{2}$.
	In addition to their fundamental relevance, our findings have important practical consequences for optimal resource management in quantum technologies.
	
\end{abstract}

\maketitle

According to von Neumann's prescription~\cite{Book_Jordan2024quantum}, a quantum measurement starts with a \textit{pre-measurement}, an interaction between a target system and a second one called a \textit{quantum meter}. 
The latter is then collapsed by a classical measurement apparatus. For example, in a Stern-Gerlach experiment, the particles' spin is measured by mapping it onto their position that is then observed on a detector screen. The Wigner-Araki-Yanase (WAY) theorem~\cite{Busch2010translation,Araki1960,Yanase1961Optimal} and its generalizations~\cite{Ozawa2002Conservation,Luo2003Wigner,Miyadera2006WAYtheorem,Loveridge2011Measurement,Navascues2014HowEnergy,Tajima2018Uncertainty,Tajima2019OzawaBoundSaturation,Tajima2020CoherenceCost,Kuramochi2023WAYUnbounded,Tajima2022Universal,Mohammady2023measurement} pose limits on the accuracy of quantum measurements due to the presence of globally conserved quantities between system and meter. In particular, Ozawa derived an expression for their minimal inaccuracy, often dubbed Ozawa's bound~\cite{Ozawa2002Conservation}. On a practical level, these bounds are inconsequential when the meter is a macroscopic apparatus~\cite{Ozawa2002Conservation}. However, they are now gaining practical importance as error correction~\cite{Book_Nielsen2010,Ziqian2024Autonomous_Experiment} and quantum sensors~\cite{Devoret_review,Blais2021CQED} rely on the speed and fidelity of quantum measurements performed using quantum meters. In this context, understanding fundamental limits on measurements’ accuracy is increasingly critical; it prevents researchers from engineering against physical laws, and guides the design of quantum architectures towards the saturation of these bounds.

One natural scenario for the application of the WAY theorem and Ozawa's bound can be found within scattering-type measurements~\cite{Book_Taylor2006Scattering,Katsube2023limitations}. In such dynamics, the interaction between meter and system is negligible at the initial and final times so that the globally conserved quantity is the sum of the bare Hamiltonians of system and meter. The effective duration of the pre-measurement is given by the time interval where system's and meter's wavefunctions overlap. Such scattering-type measurements allow for an initialization of the meter at an arbitrary time in the past and a readout at an arbitrary time in the future. In this scenario, limitations to the accuracy of quantum measurements arise whenever the system's observable does not commute with its bare Hamiltonian, thus being non-stationary and not allowing for so-called quantum nondemolition measurements~\cite{Book_Haroche2006}.
Such measurement accuracy is limited by Ozawa's bound, which in this scenario entails energetic features of the system and the meter. Namely, high-accuracy measurements require the meter's energy dispersion to be much larger than the system's energy scale, making the change in energy of the meter hardly detectable. 
Notably, the system's energy change due to the measurement can be exploited to build quantum engines~\cite{Bresque2021Two-Qubit}, thus being named \textit{quantum heat}~\cite{Elouard2017Role,Bresque2021Two-Qubit,Linpeng2024QuantumEnergetics} or \textit{measurement energy}~\cite{Rogers2022Postselection}.

In this letter, we consider such a scattering-type measurement based on a feasible interferometric scheme.
In the considered setting, the quantum meter is a flying particle described by the so-called quantum clock Hamiltonian~\cite{Aharonov1984Quantum,Aharonov1998Measurement,Malabarba2015clock,Gisin2018Quantum,Soltan2021Conservation,Spencer2022Postselection,Piccione2024ReservoirFree}, and the target system is a qubit. The clock Hamiltonian greatly simplifies our treatment by describing a wavepacket which moves at constant speed, without spatial deformations, and without back-scattering.
The interferometer is adjusted to map a qubit observable on the position of the flying particle. In this setting, errors arise as soon as the observable is not stationary and we compute their quantification $\varepsilon$ explicitly.
Denoting by $\Delta t$ the meter-target effective interaction time and by $\omega_q$ the qubit's frequency, we find that $\varepsilon \propto \Delta t$ when $\omega_q \Delta t \ll 1$.
Comparing $\varepsilon$ to the lower-bound error given by Ozawa, $\varepsilon_{\mathrm B}$, we can reformulate their ratio into an intuitive time-energy \enquote{uncertainty relation} $\Delta t \Delta E = (\hbar/2) (\varepsilon/\varepsilon_{\mathrm B}) \geq \hbar/2$, where $\Delta E$ is the meter's energy dispersion.
As a consequence, we find that the bound can be (approximately) saturated if and only if the meter is prepared in a Gaussian wavepacket, the minimum uncertainty state.
In the opposite regime of long interaction times ($\omega_q \Delta t \gg 1$), we get instead $\varepsilon = \sqrt{2} \varepsilon_{\mathrm B}$, irrespective of wavepacket's shape.
To the best of our knowledge, the implementation of our scheme or a similar one would provide the first experimental test of Ozawa’s bound. This test can arguably be implemented within state-of-the-art architectures of waveguide quantum photonics, superconducting circuits, and atomic physics.

\vskip 2mm
\emph{Ozawa's bound -- a quick summary}:
Let us consider a target system $S$ interacting with a meter system $M$ within a joint unitary dynamics described by the unitary operator $U$. We will denote by $O_S$ the observable we wish to measure on $S$, and by $O_M$ the pointer observable. After the pre-measurement, the latter is measured by a classical measurement apparatus to perform the readout of $O_S$. 
Finally, let $L_S\otimes \Id_M$ and $\Id_S \otimes L_M$ be two Hermitian operators acting, respectively, only on $S$ and $M$ such that their sum is a conserved quantity. Hereafter, we avoid writing identities in tensor products when this does not create confusion so that, for example, we write $\comm{L_S + L_M}{U} =0$. Following Ref.~\cite{Ozawa2002Conservation}, we define a noise operator $N = U^\dg O_M U - O_S$ giving the readout error (or measurement's inaccuracy) 
\begin{equation}\label{eq:ErrorOzawaFormula}
	\varepsilon^2 (\ket{\psi_S}) = \ev{N^2}{\psi_S, \psi_M},
\end{equation}
where $\ket{\psi_S}$ and $\ket{\psi_M}$ are, respectively, the states of system and meter before the interaction $(t=t_0)$.
Assuming $\comm{L_M}{O_M}=0$, (the so-called Yanase's condition~\cite{Loveridge2011Measurement}) Ozawa derived the following bound for the error:
\begin{equation}
	\label{eq:OzawaBound}
	\varepsilon^2 (\ket{\psi_S}) \geq \tilde{\varepsilon}^2_{\mathrm B} (\ket{\psi_S})
	=  \frac{1}{4} \frac{\abs{\ev{\comm{O_S}{L_S}}{\psi_S}}^2}{\Delta L_S^2 + \Delta L_M^2},
\end{equation}
where the variance $\Delta L_S^2 = \ev{L_S^2} - \ev{L_S}^2$ is computed on the initial state of $S$ and similarly for $\Delta L_M^2$. Remarkably, the bound does not depend on the interaction between system and meter nor on the pointer observable $O_M$. The initial state of the meter $M$ only enters the bound through the variance $\Delta L_M$. In agreement with the WAY statement, the bound goes to zero when $\Delta L_M \rightarrow \infty$. Hence, for all practical purposes, Ozawa's bound is negligible when macroscopic meters are employed. 
Notably, the asymptotic saturation of this bound for $\tilde{\varepsilon}_{\mathrm B} \rightarrow 0$ has been shown in Ref.~\cite{Tajima2019OzawaBoundSaturation}. Moreover, it has been improved in various ways for mixed states of system and meter~\cite{Tajima2019OzawaBoundSaturation} and generalized measurement schemes~\cite{Mohammady2023measurement}. Here, however, we will consider pure states of meter and system as well as standard observables represented by Hermitian operators, where these generalizations reduce to the original Ozawa's bound.

\begin{figure}[t]
	\centering
	\includegraphics[width=0.48\textwidth]{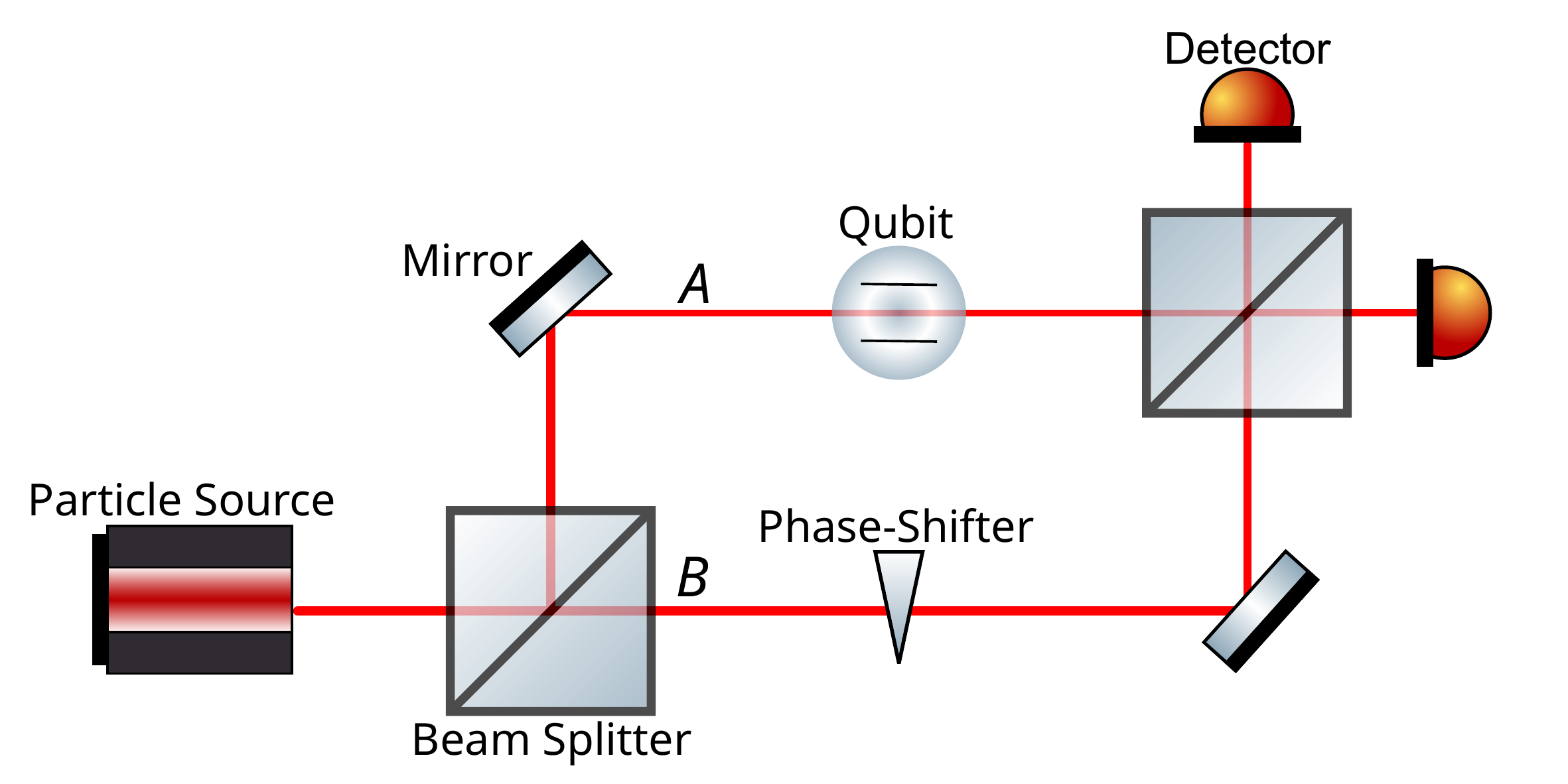}
	\caption{Setup of the proposed measurement scheme, based on a Mach-Zehnder interferometer. A flying particle (the meter) passes through a balanced beamsplitter. In path $A$, the particle interacts with the qubit, in path $B$, it always acquires the same phase $\pi/2$. The two paths then recombine by means of a second balanced beamsplitter. In the case of an ideal projective measurement of $\sigma_z$, corresponding to $\prtg{\Theta,\phi}=\prtg{0,\pi}$, the particle is revealed with certainty by the upper detector if the qubit state is $\ket{g_z}$ and by the other one if it is $\ket{e_z}$.}
	\label{fig:ParticleInterferenceSetup}
\end{figure}

\vskip 2mm
\emph{Measurement setup:}
We consider a measurement scheme where $S$ is a qubit and $M$ is a flying particle. The pre-measurement is performed within a Mach-Zehnder interferometer, while the classical measurement of $M$ takes place at the output arms of the interferometer, as depicted in Fig.~\ref{fig:ParticleInterferenceSetup}. 
We denote the eigenstates of the qubit along the $z$-axis of the Bloch sphere by $\ket{g_z}$ and $\ket{e_z}$ which are, respectively, the ground and excited states.
The qubit can be fixed in space (as in Fig.~\ref{fig:ParticleInterferenceSetup}) or be attached to the flying particle as an internal degree of freedom. The qubit's Hamiltonian is $H_S = (\hbar \omega_q/2)\prtq{\cos(\Theta) \sigma_z + \sin(\Theta) \sigma_x}$ where $\omega_q$ is its bare frequency, $\Theta$ is an angle that can vary in $[0,\pi/2]$, and $\sigma_x = \dyad{g_z}{e_z} + \dyad{e_z}{g_z}$ and $\sigma_z = -\dyad{g_z} + \dyad{e_z}$ are Pauli matrices. 

We take the Hamiltonian of the flying particle to be the quantum clock Hamiltonian~\cite{Aharonov1984Quantum,Aharonov1998Measurement,Malabarba2015clock,Gisin2018Quantum,Soltan2021Conservation,Spencer2022Postselection,Piccione2024ReservoirFree}, $H_M = v_0 \hp$. 
It can be seen as the Hamiltonian of a single photon propagating in a one-dimensional path with a linear dispersion relation~\cite{Shen2009Theory,Piccione2024Fundamental}. It can also be obtained as the linearization of kinetic energy in solid-state physics~\cite{Yamamoto2012Electrical} or of the free particle Hamiltonian $\hp^2/2m$~\cite{Piccione2024ReservoirFree}. 
This latter approximation can be made arbitrarily good by increasing the mass $m$ of the flying particle while keeping its average speed constant~\cite{Piccione2024ReservoirFree}.
Finally, the Hamiltonian governing the system-meter dynamics is
\begin{equation}
	\label{eq:Hamiltonian}
	H = H_M + H_S + V(\hx),
	\quad
	V(\hx)=\frac{\hbar \phi v_0}{2}\delta_A (\hx) \otimes \sigma_z,
\end{equation}
$\phi \in [0,2\pi]$ gives the strength of the pre-measurement. In the commuting case where $H_S \propto \sigma_z$, $\phi/2$ is the phase shift acquired by the flying particle as a result of the interaction with the qubit.
The Dirac delta function $\delta_A (\hx)$ indicates that the scattering interaction happens at position $x=0$ in the arm $A$ of the interferometer and allows for analytical solutions. A more general interaction term could be $V(\hx)= \hbar\sigma_z f_A (\hx)$, with $f_A(x)$ characterized by a typical length $L$. However, when $L \ll v_0/\omega_q$, the clock Hamiltonian implies (see Appendix~\ref{APPSec:ScatteringMap}) that the dynamics are well approximated by the interaction Hamiltonian $V(\hx)=\hbar\prtq{\int f_A (x) \dd{x}}\delta_A (\hx)v_0 \sigma_z$, thus justifying our use of the Dirac delta in the interaction term. We will comment more on the experimental feasibility of the Hamiltonian~\eqref{eq:Hamiltonian} in the last section.

At $t=t_0<0$, prior to the beginning of the interaction, the meter's state is described by the wavepacket $\psi_0 (x)$, with a spatial spread $\Delta x^2 \equiv \intmp (x-x_0)^2 \abs{\psi_0 (x)}^2 \dd{x}$, where $x_0=v_0 t_0$ is the wavepacket's average position at $t_0$, and $v_0$ is its group velocity. Since we consider the qubit as a point-like entity, the effective duration ($\Delta t$) of the pre-measurement is perfectly determined by the wavepacket spatial spread, i.e., $\Delta t = \Delta x/v_0$. 
Moreover, due to the clock Hamiltonian, the particle's frequency and momentum are linked by $\omega=(v_0/\hbar)p$. Then, we can rewrite Heisenberg's uncertainty relation $\Delta p \Delta x \geq \hbar/2$ for the meter's wavepacket as $\Delta \omega \Delta t \geq 1/2$. We stress that $\Delta t$ has the physical meaning of \enquote{effective interaction time} of the pre-measurement but it does not come from a time operator.

The setup in Fig.~\ref{fig:ParticleInterferenceSetup} measures the qubit observable $O_S=\sigma_z$ as follows. First, the flying particle goes through a first beam splitter which generates a balanced superposition of the output paths $A$ and $B$. 
More precisely, the transformations implemented by the beam splitters are $\ket{\psi}_A\ket{0}_B \rightarrow [1/\sqrt{2}]\prtq{\ket{\psi}_A\ket{0}_B+\ket{0}_A\ket{\psi}_B}$ and $\ket{0}_A\ket{\psi}_B \rightarrow [1/\sqrt{2}]\prtq{\ket{\psi}_A\ket{0}_B - \ket{0}_A\ket{\psi}_B}$, where $\ket{\psi}$ is an arbitrary one-particle state while $\ket{0}$ is the vacuum state. We assume that the beam splitters have the same response for all relevant frequencies of the traveling wavepacket. 
Then, the interaction with the target qubit takes place in the arm $A$, while the arm $B$ contains a phase shifter introducing a phase of $\pi/2$. Finally, the second balanced beam-splitter recombines the two paths and then a classical measurement finds out whether the particle is in path $A$ or $B$, performing the qubit's readout.
This final measurement is implemented by perfectly efficient time-resolved detectors positioned at each output arm of the interferometer. These detectors provoke the instantaneous collapse of the meter's wavepacket at the detector's position, and their measurement is integrated over a long time with respect to the wavepacket's duration. 

Let us illustrate the working principle of the setting in the simplest case, where it achieves an ideal and quantum non-demolition projective measurement of $\sigma_z$. This case corresponds to taking two assumptions: $\Theta = 0$, and $\phi=\pi$. The latter assumption is known in quantum optics as the $\pi$-per-photon condition~\cite{Book_Haroche2006}.
Under these assumptions, the flying particle acquires a factor $e^{\pm i \pi/ 2}$ depending on the qubit state along $\sigma_z$.
At the output of the second balanced beam splitter, the particle is found in path $A$ of the interferometer if the qubit was in $\ket{e_z}$, and in path $B$ of the interferometer if the qubit was in $\ket{g_z}$. Hence we define the pointer observable as $O_M = \hn_A - \hn_B$, where $\hn_A$ measures if the particle is in $A$ with outcomes $0$ (no particle) and $1$ (there is a particle) and $\hn_B$ does the same for path $B$. The WAY theorem's assumptions are satisfied upon the identification $L_S = H_S$ and $L_M = H_M$ because beamsplitters are energy conserving at the scattering level~\footnote{In other words, the energy of the particle before and after the interaction with the beam-splitter is the same.} and the particle-qubit interaction is of the scattering type~\cite{Book_Taylor2006Scattering} thus implying $\comm{H_S+H_M}{U}=0$, where the unitary operator $U$ describes the entire pre-measurement dynamics (beamsplitters included).

\vskip 2mm
\emph{Readout error}:
Now, we consider the case where the qubit's Hamiltonian $H_S$ is tilted by an angle $\Theta \in (0,\pi/2]$ so that $\comm{H_S}{\sigma_z}\neq 0$. Hence, the observable $\sigma_z$ one wishes to measure is not stationary and the measurement cannot be quantum non-demolition.
Within an ideal projective measurement scheme, we would have to specify the time at which the measurement is implemented. Here, since the pre-measurement takes a finite amount of time $\Delta t$, we aim to measure $\sigma_z$ as it was at the time $t=t_0$ prior to the beginning of the interaction.
Hereafter, we also relax the condition on the phase acquired by the flying particle, considering also the cases where the angle $\phi$ is different from $\pi$. However, we will see that the best accuracy is always attained for $\phi=\pi$.

Let us write the qubit's state at $t=t_0$ as $\ket{\psi_S} = b_g \ket{g_\Theta} + b_e \ket{e_\Theta}$, where $\ket{g_\Theta}$ and $\ket{e_\Theta}$ are, respectively, the ground and excited states of $H_S$. In Appendix~\ref{APPSec:ScatteringMap}, we show that the interaction between the particle and the qubit (in the interaction picture) leads to the map:
\begin{multline}
	\label{eq:ScatteringMap}
	\ket{\psi_S, 1_\omega} \rightarrow 
	b_g I_{gg} \ket{g_\Theta,1_\omega} + b_e I_{ee} \ket{e_\Theta,1_\omega} +\\+
	b_g I_{ge}\ket{e_\Theta,1_{\omega-\omega_q}} + b_e I_{eg}\ket{g_\Theta,1_{\omega + \omega_q}},
\end{multline}
where $I_{gg} = I_{ee}^* = \cos(\phi/2)+ i \cos(\Theta)\sin(\phi/2)$, $I_{ge} = I_{eg} = i \sin(\Theta)\sin(\phi/2)$, and $\ket{1_\omega}$ is the initial meter's state, whose average frequency is $\omega$.
The wavefunctions $\ket{1_{\omega \pm \omega_q}}$ are spatially shaped as the input one but shifted in frequency by the qubit's frequency $\omega_q$, as expected from energy conservation in scattering processes~\cite{Piccione2024Fundamental}.
The readout error defined by Eq.~\eqref{eq:ErrorOzawaFormula} then reads (see Appendix~\ref{APPSec:CalculationsError}):
\begin{equation}
	\label{eq:ErrorFormula}
	\varepsilon^2 = 
	2\prtg{1 - \sin(\frac{\phi}{2})\prtq{\cos[2](\Theta)+P\sin[2](\Theta)}},
\end{equation}
where $P$ stands for the overlap between the initial wavepacket and a frequency-shifted one, i.e.,
\begin{equation}
	\label{eq:OverlapShift}
	P 
	\equiv \Re{\ip{1_\omega}{1_{\omega + \omega_q}}}
	\!=\!
	\int \dd{x} \abs{\psi_0 (x)}^2 \cos(\frac{\omega_q (x-x_0)}{v_0}).
\end{equation}
We can see that the highest accuracy is attained for $\phi=\pi$ for any value of $\Theta$. Setting $\phi=\pi$ we can write
\begin{equation}\label{eq:ErrorFormulaOptimized}
	\varepsilon=\sin(\Theta)\sqrt{2(1-P)}.
\end{equation}
As expected, the above equation reveals that perfect accuracy ($\varepsilon=0$) is attained in the full commuting case, $\Theta=0$, for any meter wavepacket. Alternatively, for $\Theta\neq 0$, it can still be attained when $P \rightarrow 1$, which entails that the shifted meter's wavepackets are completely indistinguishable from the input one, albeit being shifted in frequency by $\pm \omega_q$ [cf. Eq.~\eqref{eq:ScatteringMap}].
Hereafter, we refer to this regime as the good measurement regime.
Of course, this regime is attainable only when $\Delta \omega \gg \omega_q$. In this regime, the meter does not extract information on the qubit's bare energy $H_S$ but rather about the target observable $\sigma_z$.
From a temporal viewpoint, $P \rightarrow 1$ captures
the regime where $\Delta t \ll \omega_q^{-1}$, meaning that the qubit evolution can be considered as frozen during the pre-measurement interaction. 
In the opposite regime, a pre-measurement with $P<1$, i.e., characterized by a finite duration, is akin to a time-averaged (over a time $\Delta t$) measurement of $\sigma_z$, which evolves due to the qubit's bare dynamics.
From a spectral view point, this is the regime where the meter carries information on the qubit energy state. The meter's energy shifts can be used to access quantum heat exchanges, as done in Ref.~\cite{Linpeng2024QuantumEnergetics}.

\vskip 2mm
\emph{Relation with Ozawa's bound:}
We now compare the readout error studied above to Ozawa's bound [Eq.~\eqref{eq:OzawaBound}], expressed in our setting.
As the error given in Eq.~\eqref{eq:ErrorFormulaOptimized} is independent of the initial qubit's state, we compute Ozawa's bound by maximizing over all possible initial qubit states (see Appendix~\ref{APPSec:BoundMaximization}), obtaining
\begin{equation}\label{eq:OzawaBoundQubit}
	\varepsilon_{\mathrm B}^2 = \max_{\ket{\psi_S}} \tilde{\varepsilon}_{\mathrm B}^2 \prt{\ket{\psi_S}} =
	\frac{\sin[2](\Theta)}{1 + 4 (\Delta \omega/\omega_q)^2}.
\end{equation}
The bound only depends on the energy dispersion of the meter compared to the qubit's energy scale. 
We stress that $\Delta \omega \gg \omega_q$ is a necessary condition for an accurate measurement but not a sufficient one. It is in fact easy to consider meter's wavefunctions with both $\Delta \omega \gg \omega_q$ and $\Delta t \gg \omega_q^{-1}$.
In other words, meter's wavefunctions for which the measurement's accuracy is poor despite Ozawa's bound allowing for a very high accuracy measurement.

First, we analyze the good measurement regime characterized by $\omega_q \Delta t \ll 1$, which implies $\Delta \omega \gg \omega_q$ due to Heisenberg's uncertainty principle ($\Delta \omega \Delta t \geq 1/2$). In other words, the interaction time between meter and qubit is much shorter than $1/\omega_q$. This is the regime in which, despite having $\Theta>0$, accurate measurements of $\sigma_z$ can be attained and thus the one of most practical importance. 
We notice that Ozawa's bound [Eq.~\eqref{eq:OzawaBoundQubit}] can be approximated to $\varepsilon_{\mathrm B} \simeq \sin(\Theta) \omega_q/ (2 \Delta \omega)$ and Eq.~\eqref{eq:OverlapShift} to $P \simeq 1-(1/2)\omega_q^2 \Delta t^2$ such that the error reads [cf. Eq.~\eqref{eq:ErrorFormulaOptimized}] $\varepsilon \simeq \sin(\Theta)\omega_q \Delta t$.
Then, we can write
\begin{equation}\label{eq:FrequencyTimeUncertainty}
	\Delta \omega \Delta t \simeq \frac{1}{2}\frac{\varepsilon}{\varepsilon_{\mathrm B}} \geq \frac{1}{2},
\end{equation}
which entails $\Delta \omega \Delta t = 1/2 \implies \varepsilon \simeq\varepsilon_{\mathrm {B}}$.
Hence, in the good measurement regime, Ozawa's bound can be practically saturated, but only by Gaussian wavepackets as they are the only wavepackets satisfying the minimum uncertainty relation~\cite{Book_Cohen2019QuantumMechanicsVol1}.

\begin{figure}[t]
	\centering
	\includegraphics[width=0.48\textwidth]{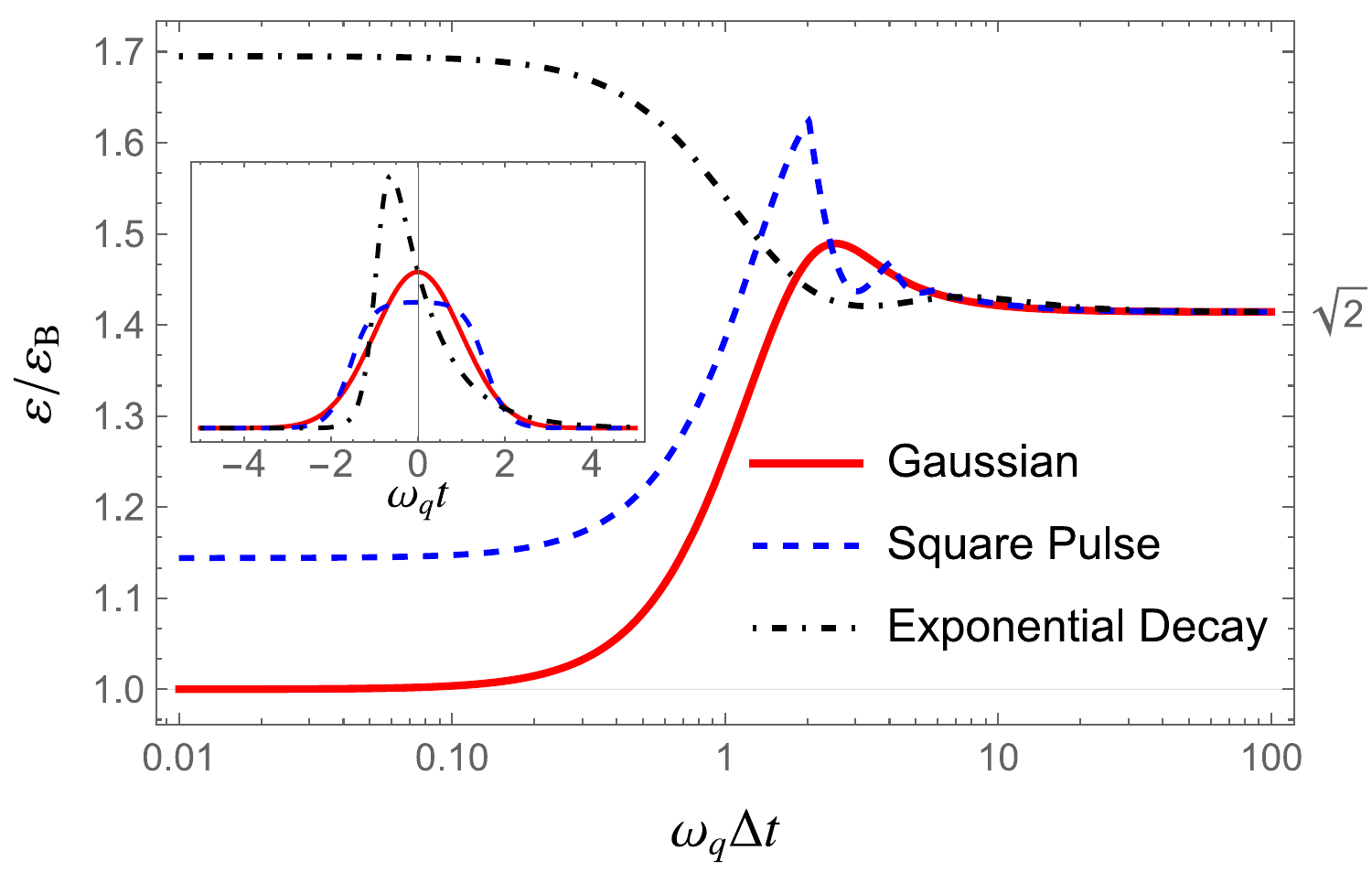}
	\caption{Ratio $\varepsilon/\varepsilon_{\mathrm B}$ as a function of $\omega_q \Delta t$ for different wavepacket shapes. The inset in the top-left corner shows the three shapes: Gaussian, square pulse, exponential decay.
		For each given wavepacket shape, we can compute $\Delta \omega$ as a function of $\Delta t$ and use it to compute $\varepsilon_{\mathrm B}$ (see Appendix~\ref{APPSec:WavepacketShapesErrorBounds}).
		As predicted, only the Gaussian wavepacket saturates Ozawa's bound for $\omega_q \Delta t \ll 1$. In the vicinity of $\omega_q \Delta t = 1$, wave packet shapes other than the Gaussian can have a higher accuracy. Moreover, for every shape, we get $\varepsilon/\varepsilon_{\mathrm B}=\sqrt{2}$ for $\omega_q \Delta t \gg 1$, as predicted.}
	\label{fig:ErrorRatioPlotsInset}
\end{figure}

For longer interaction times, the bound is no longer saturated. However, another general result can be obtained in the regime $\Delta \omega \ll \omega_q$ (which implies $\omega_q \Delta t \gg 1$ and $P \rightarrow 0$): the ratio between error and Ozawa's bound gives $\varepsilon/\tilde{\varepsilon}_{\mathrm B} = \sqrt{2}$ independently of the meter's wavepacket shape.
Here, the pre-measurement can be seen as some sort of time-averaged measurement of the observable $\sigma_z$. Its eigenstates rotate around an axis forming an angle $\Theta$ with the $z$-axis of the Bloch sphere due to the qubit's free evolution. In particular, when $\Theta=\pi/2$, these eigenstates rotate around the $x$-axis so that the measurement cannot extract any information on the value of $\sigma_z$ at $t=t_0$. Indeed, for $P \rightarrow 0$ and $\Theta=\pi/2$ the error $\varepsilon$ of Eq.~\eqref{eq:ErrorFormula} takes the same value independently of $\phi$, including $\phi=0$ (i.e., in the absence of interaction).

In Fig.~\ref{fig:ErrorRatioPlotsInset} we display the ratio $\varepsilon/\varepsilon_{\mathrm B}$ as a function of $\omega_q \Delta t$ for different wavepacket shapes. We can see that, indeed, only the Gaussian wavepacket saturates Ozawa's bound. 
The comparison is particularly interesting in the good measurement regime ($\omega_q \Delta t \ll 1$), the most relevant one for applications. The plot shows that exponentially decreasing wavepackets, often taken as the natural choice of single-photon wavepackets in waveguide QED, lead to an error that is $70\%$ higher compared to Gaussian wavepackets.
More details on how these plots are obtained can be found in Appendix~\ref{APPSec:WavepacketShapesErrorBounds}.

\vskip 2mm
\emph{Possible experimental implementations:}
The setup of Fig.~\ref{fig:ParticleInterferenceSetup}, or an equivalent one, can be implemented in various physical platforms. Options include circuit QED, quantum photonics, and flying particles with internal degrees of freedom (IDoF). Among these possibilities, the most-promising one appears to us the one implementable in circuit QED settings. We will first briefly comment on the last two and then, more in detail, on the first one.

Regarding flying particles, our scheme could be implemented by treating their IDoF~\cite{Piccione2024ReservoirFree} as the qubit to measure. A spatially localized field acting on the IDoF can implement the system-meter interaction. Moreover, if there is no energy difference between the IDoF states, a global field can be used to implement $H_S$. 
In this case, for the clock Hamiltonian ($H_M=v_0 \hp$) to provide a good description of the dynamics, the condition $\ev{\hp^2/2 m} \gg \ev{H_S+V(\hx)}$ has to be satisfied at all times for all possible qubit states. One needs to assume that the kinetic energy changes induced by $V(\hx)$ are negligible at all times~\cite{Piccione2024ReservoirFree}.
Finally, notice that beamsplitters for massive particles such as electrons~\cite{Ji2003ElectronicMachZehnder}, atoms~\cite{Huesmann1999SingleAtom,Cronin2009Interferometry,Machluf2013CoherentSternGerlach,Barrett2014TheSagnac,Keil2016Fifteen,Amico2021RoadmapAtomtronics,Margalit2021Realization}, and molecules~\cite{Cronin2009Interferometry,Keil2016Fifteen,Amico2021RoadmapAtomtronics} are actively investigated.

In quantum photonics, a single-photon wavepacket can play the meter role and the qubit states could be encoded in two possible propagation paths. By making these paths close enough, the photon can hop among them during the propagation, thus implementing an Hamiltonian $H_S \propto\sigma_x$~\cite{Peruzzo2010QuantumWalks,AspuruGuzik2012PhotonicReview,Wang2020PhotonicReview} whose magnitude can be much smaller than the photon energy (as in Ref.~\cite{Peruzzo2010QuantumWalks}) while the scattering-type interaction $V(\hx)\simeq \delta (\hx) \otimes \sigma_z$ of Eq.~\eqref{eq:Hamiltonian} could be implemented by inserting two different refractive materials in the two paths.

In circuit QED, dispersive interactions followed by interferometric readouts are widely used to measure superconducting (transmon) qubits in microwave resonators~\cite{Devoret_review,Blais2021CQED}. The Hamiltonian $H_S$ with $\Theta \neq 0$ can be implemented by driving the qubit with a classical field~\cite{Blais2004Architecture,Blais2007Processing,Gambetta2008Zeno,Boissonneault2009Dispersive,Blais2021CQED,Linpeng2024QuantumEnergetics} while the meter Hamiltonian can be obtained by considering a single photon traveling in the waveguide~\cite{Shen2009Theory,Piccione2024Fundamental}. In Ref.~\cite{Linpeng2024QuantumEnergetics}, a qubit Hamiltonian with $\Theta=\pi/2$ has been implemented in this way, with a frequency $\omega_q/2\pi \sim 1\ \text{MHz}$. The natural qubit Hamiltonian along the $z$-axis of the Bloch sphere, with frequency $\omega_z/2\pi \sim \text{GHz}$, is eliminated by studying the dynamics in the rotating frame with the same frequency. In this case, the central frequency of the traveling photon is around $\omega/2\pi \sim 5\ \text{GHz}$ so that its frequency dispersion can be $\Delta \omega/2\pi \sim 10\ \text{MHz}$ achieving $\omega \gg \Delta \omega \gg \omega_q$, thus making the good-measurement regime physically allowed. 
The experimental measurement of the readout error as defined by Ozawa requires separate access to the measurements of $O_M$ and $O_S$, which is easily achievable in this kind of platform. 
Finally, pulse-shaping techniques can be used to make various types of one-photon wavepackets as in Fig.~\ref{fig:ErrorRatioPlotsInset}.

\vskip 2mm
\emph{Conclusions and outlook:}
We proposed an interferometric measurement scheme allowing the investigation of fundamental measurement limitations due to conservation laws.
In the good measurement regime, we mapped Ozawa's bound into an energy-time uncertainty relation of the flying meter’s wavepacket. We then showed that the bound can be approximately saturated if and only if the flying meter is prepared in a Gaussian wavepacket, the one of minimum time-energy uncertainty.
Finally, we argued how our scheme could be experimentally implemented, allowing for a first experimental test of Ozawa's bound, in the same vein of recent experimental progress pushing the experimental verification of fundamental measurement bounds (see for example~\cite{Hong2022InformationTradeOff}).

The accuracy bound explored in this letter as well as other fundamental bounds investigated in the literature have practical consequences for quantum technologies. Although macroscopic meters make them negligible, this is not so when quantum meters (sometimes called ancillas) are used~\cite{Ozawa2002ConservativeQuantumComputing,Katsube2023limitations}. In fact, while increasing the size of the meter can increase the accuracy of the measurement, it can also greatly increase the practical energetic cost of the measurement operation~\cite{Fellous-Asiani2021Limitations,FellousAsiani2023Optimizing}. Therefore, measurement schemes saturating these bounds can be of great importance for the energetic cost of quantum technologies~\cite{Auffeves2022QEI}.

\vskip 1mm
\emph{Acknowledgements.---} This work was supported by the John Templeton Foundation (Grant No. 61835), the Foundational Questions Institute Fund (Grants No. FQXi-IAF19-01 and FQXi-IAF19-05), the ANR Research Collaborative Project ``QuRes" (Grant ANR-PRC-CES47-0019) and  “Qu-DICE” (Grant ANR-PRC-CES47), the National Research Foundation, Singapore and A*STAR under its CQT Bridging Grant, the Plan France 2030 through the project NISQ2LSQ (Grant ANR-22-PETQ-0006) and the project OQuLus (Grant ANR-23-PETQ-0013), the project BACQ (Grant ANR-22-QMET-0002). M. M. acknowledges the support by PNRR MUR project PE0000023-NQSTI. N. P. acknowledges the support by PNRR NQSTI Spoke 1 “Foundations and architectures for quantum information processing and communication” CUP: J93C22001510006.

We thank M. Richard and R. Whitney for useful discussions about possible experiments to verify our results. N.P. thanks R. Takagi and H. Wilming, for useful discussions about the results of this work.


%

\clearpage
\onecolumngrid
\appendix

\section{Scattering map\label{APPSec:ScatteringMap}}
In this section, we derive Eq.~\eqref{eq:ScatteringMap} of the main text. The bipartite system we study is composed of the following Hamiltonian:
\begin{equation}
	H= H_S + H_M + V,
	\qquad
	H_S = \frac{\hbar \omega_q}{2}\prtq{\cos(\Theta)\sigma_z + \sin(\Theta)\sigma_x},
	\quad
	H_M = v_0 \hq,
	\quad
	V = \frac{\hbar \phi v_0}{2}f(\hx)\sigma_z,
\end{equation}
where $f(\hx)$ is a generic function of the position operator $\hx$ normalized so that $\int f(x) \dd{x}=1$ and characterized by a length $L$ such that $\abs{x}\gg L \implies f(x)\sim 0$. As in the main text, the initial state of the qubit is $\ket{\psi_S}=b_g\ket{g_\Theta} + b_e \ket{e_\Theta}$. The initial state of the meter is instead given by $\ket{1_\omega}$, representing a wavepacket centered around the frequency $\omega$. The frequency, here, corresponds to the wavevector multiplied by $v_0$, i.e., $\omega = v_0 k$, where $k = p/\hbar$ is a wavevector. Notice that, with respect to the main text, here the dynamics starts at $t = 0$ to lighten the notation. In other words, at $t=0$, the meter's wavepacket is still far on the left of the qubit's position ($x=0$).

Let us denote the initial wavefunction of the meter by $\psi_M (x)$. It can be seen\footnote{See the Supplemental Material of Ref.~\cite{Piccione2024ReservoirFree} for more details.} that 
\begin{equation}
	\ket{\psi(t)} = \intmp \dd{x} \psi_M (x-v_0 t) U(x+v_0 t; x) \ket{x}\ket{\psi_S},
	\quad
	U(x+v_0 t; x) = \mcT\exp{-\frac{i}{\hbar}\int_{0}^{t} \prtq{H_S + \frac{\hbar \phi v_0}{2}f(x + v_0 s)\sigma_z} \dd{s}},
\end{equation}
by inserting the above formula into the \Schr equation. 
Let us now consider the regime where $\omega_q \ll v_0/L$. The bare dynamics of the qubit is then practically frozen during a timescale $L/v_0$ and we can make the substitution $f(x) \rightarrow \delta (x)$. We also consider the long-time limit for which $t$ is always high enough such that $s =-x/v_0$ is always within the integration extremes.
Then, the unitary operator can be written as follows:
\begin{equation}
	U(x+v_0 t; x) = e^{-(i/\hbar)(t+x/v_0)H_S}e^{-i (\phi/2)\sigma_z}e^{-(i/\hbar)(-x/v_0)H_S}.
\end{equation}
Going to interaction picture with respect to both system and meter Hamiltonians, the state at time $t$ is
\begin{equation}
	\ket{\psi_I (t)} =\exp[\frac{i}{\hbar}(H_S+H_M)t] \ket{\psi (t)}= \intmp \dd{x} \psi_M (x) e^{(i/\hbar)(-x/v_0)H_S}e^{-i (\phi/2) \sigma_z}e^{-(i/\hbar)(-x/v_0)H_S} \ket{x}\ket{\psi_S}.
\end{equation}
By substituting $\ket{\psi_S}$ with its decomposition in the $H_S$ basis and inserting the identity term $\dyad{g_\Theta}+\dyad{e_\Theta}$ where needed one arrives at 
\begin{equation}
	\ket{\psi_S, 1_\omega} \rightarrow 
	b_g I_{gg} \ket{g_\Theta,1_\omega} + b_e I_{ee} \ket{e_\Theta,1_\omega} +
	b_g I_{ge}\ket{e_\Theta,1_{\omega-\omega_q}} + b_e I_{eg}\ket{g_\Theta,1_{\omega + \omega_q}},
\end{equation}
where 
\begin{equations}
	\label{APPeq:JumpConstants}
	I_{gg} &= I_{ee}^* = \ev{e^{-i (\phi/2) \sigma_z}}{g_\Theta}=\cos(\phi/2)+ i \cos(\Theta)\sin(\phi/2),\\
	I_{ge} &= I_{eg} =\mel{g_\Theta}{e^{-i (\phi/2) \sigma_z}}{e_\Theta}= i \sin(\Theta)\sin(\phi/2),
\end{equations}
and we have
\begin{equations}
	\ket{1_\omega} &= \intmp \dd{x} \psi_M (x) \ket{x} = \intmp \dd{p} \tpsi_M (p) \ket{p}, \qq{where} \tpsi(p)=\frac{1}{\sqrt{2\pi \hbar}}\intmp \dd{x} \psi_M (x) e^{-i x p/\hbar},\\
	\ket{1_{\omega+\omega_q}} &= \intmp \dd{x} \psi_M (x) e^{+ i\frac{\omega_q}{v_0}x}\ket{x}
	= \intmp \dd{p} \tpsi_M \prt{p-\frac{\hbar \omega_q}{v_0}} \ket{p},\\
	\ket{1_{\omega-\omega_q}} &= \intmp \dd{x} \psi_M (x) e^{- i\frac{\omega_q}{v_0}x}\ket{x} 
	= \intmp \dd{p} \tpsi_M \prt{p + \frac{\hbar \omega_q}{v_0}} \ket{p}.
\end{equations}
The above equations imply the frequency shifts reported in the main text as one can write $p=(\hbar/v_0)\omega$.

\clearpage

\section{Computation of Ozawa's error quantifier\label{APPSec:CalculationsError}}

Here, we compute the error of the measurement associated to the setup described in the main text. To do this we have to compute the quantity $\varepsilon^2 (\ket{\psi_S}) = \ev{N^2}{\psi_S,\psi_M}$ defined in the main text, where $N = U^\dg O_M U - O_S$. We also recall that $O_S = \sigma_z$ is the system observable we want to measure and $O_M = \hn_A - \hn_B$, which is also an observable, represents the measurement we perform on the meter to perform an indirect measurement on the system, while $U$ is the unitary operator governing the entire dynamics. Finally, $\ket{\psi_S}$ is the initial state of system $S$ and $\ket{\psi_M}$ is the initial state of the meter. Notice that, in general, this error depends on the initial state of the system we want to measure. However, we will show that this quantity is independent of the state $\ket{\psi_S}$ of the qubit in our setup.

First, since there is just one meter particle, we get that $\hn_A + \hn_B = \Id$, so that we can write that
\begin{equation}
	U N \ket{g_z,\psi_M} = 2 \hn_A U \ket{g_z,\psi_M},
	\qquad
	U N \ket{e_z,\psi_M} = - 2 \hn_B U \ket{e_z,\psi_M}.
\end{equation}
It follows that, writing $\ket{\psi_S} = c_g \ket{g_z} + c_e \ket{e_z}$, we get
\begin{equation}
	\label{APPeq:ErrorDecompositionSigmaZ}
	\varepsilon^2 (\ket{\psi_S}) 
	= \ev{N U^\dg U N}{\psi_S,\psi_M}
	= 4 \abs{c_g}^2 \ev{U^\dg \hn_A U}{g_z,\psi_M} + 4 \abs{c_e}^2 \ev{U^\dg \hn_B U}{e_z,\psi_M}.
\end{equation}
We now have to calculate these quantities.

The total output state can be computed using Eq.~\eqref{eq:ScatteringMap} of the main text\footnote{Even though Eq.~\eqref{eq:ScatteringMap} of the main text is given in interaction picture, it can be used for this calculation because $\hn_A$ and $\hn_B$ commute with the qubit's bare Hamiltonian and their value only depends on being in arm $A$ or $B$ and not on the actual position of the meter wavepacket.} and remembering the role of the two beamsplitters\footnote{We consider a beamsplitter to act as follows: $\ket{1,0}\rightarrow (1/\sqrt{2})\prtq{\ket{1,0}+\ket{0,1}}$ and $\ket{0,1} \rightarrow (1/\sqrt{2})\prtq{\ket{1,0}-\ket{0,1}}$.} and the phase shifter in arm B of the Mach-Zehnder interferometer. We get
\begin{multline}
	\label{APPeq:FinalStateDynamics}
	U\ket{\psi_S,\psi_M} =
	\frac{1}{2}\prtgB{
		\ket{g_\Theta}\prtq{b_g I_{gg}\prt{\ket{1_{\omega},0}+\ket{0,1_{\omega}}} + b_e I_{eg}\prt{\ket{1_{\omega+\omega_q},0}+\ket{0,1_{\omega+\omega_q}}} - i b_g \prt{\ket{1_{\omega},0}-\ket{0,1_{\omega}}}}
		+\\
		\ket{e_\Theta}\prtq{b_e I_{ee}\prt{\ket{1_{\omega},0}+\ket{0,1_{\omega}}}
			+ b_g I_{ge}\prt{\ket{1_{\omega-\omega_q},0}+\ket{0,1_{\omega-\omega_q}}}
			-i b_e\prt{\ket{1_{\omega},0}-\ket{0,1_{\omega}}}
		}
	},
\end{multline}
where the unitary operator $U$ is the interaction picture evolution operator. The above equations imply that (with an abuse of notation\footnote{In the following equations the particle is now localized in the arm correspondent to the applied operator, $\hn_A$ or $\hn_B$ so that we do not make explicit where the particle is for each given state as it is obvious. For example, in the equation for $\hn_B U\ket{\psi_S,\psi_M}$ the state $\ket{1_\omega}$ stands for $\ket{0,1_\omega}$.} to lighten the notation)
\begin{equations}
	\label{APPeq:FinalStatesProjected}
	\hn_A U\ket{\psi_S,\psi_M} =
	\frac{1}{2}\prtgB{
		\ket{g_\Theta}\prtq{b_g I_{gg}\ket{1_{\omega}} + b_e I_{eg}\ket{1_{\omega+\omega_q}} - i b_g \ket{1_{\omega}}}
		+
		\ket{e_\Theta}\prtq{b_e I_{ee}\ket{1_{\omega}}
			+ b_g I_{ge}\ket{1_{\omega-\omega_q}}
			-i b_e\ket{1_{\omega}}
		}
	},\\
	\hn_B U\ket{\psi_S,\psi_M} =
	\frac{1}{2}\prtgB{
		\ket{g_\Theta}\prtq{b_g I_{gg}\ket{1_{\omega}} + b_e I_{eg}\ket{1_{\omega+\omega_q}} + i b_g \ket{1_{\omega}}}
		+
		\ket{e_\Theta}\prtq{b_e I_{ee}\ket{1_{\omega}}
			+ b_g I_{ge}\ket{1_{\omega-\omega_q}}
			+ i b_e\ket{1_{\omega}}
		}
	}.
\end{equations}
We can now compute\footnote{We recall that $\hn_A= \hn_A^2$ and $\hn_B=\hn_B^2$.}
\begin{equation}
	\ev{U^\dg \hn_A U}{\psi_S,\psi_M} =
	\frac{1}{4}\prtgB{1+
		\abs{I_{gg}}^2 + \abs{I_{eg}}^2  
		+ 2 \Re\prtg{ - i \prt{\abs{b_g}^2 I_{gg}^*  + \abs{b_e}^2 I_{gg}}  
			+ 2 i I_{eg} \Re\prtg{ b_g^* b_e \ip{1_{\omega}}{1_{\omega+\omega_q}}}}
	},
\end{equation}
and
\begin{equation}
	\ev{U^\dg \hn_B U}{\psi_S,\psi_M} =
	\frac{1}{4}\prtgB{1+
		\abs{I_{gg}}^2 + \abs{I_{eg}}^2  -
		2 \Re\prtg{-i \prt{\abs{b_g}^2 I_{gg}^*  + \abs{b_e}^2 I_{gg}}  
			+ 2 i I_{eg} \Re\prtg{ b_g^* b_e \ip{1_{\omega}}{1_{\omega+\omega_q}}}}
	},
\end{equation}
where we used the fact that $\abs{b_g}^2 + \abs{b_e}^2 = 1$, $I_{gg}=I_{ee}^*$, $I_{ge} = I_{eg}$, $I_{ge}^* = - I_{ge}$, and $\ip{1_{\omega-\omega_q}}{1_\omega} = \ip{1_\omega}{1_{\omega +\omega_q}}$.

Finally, we can find the averages related to $U \ket{g_z,\psi_M}$ and $U \ket{e_z,\psi_M}$ by making the substitutions:
\begin{equations}
	\label{APPeq:BasisChange}
	\ket{\psi_S} &= \ket{g_z}  
	\qq{when} 
	b_g \rightarrow \cos(\Theta/2),\ b_e \rightarrow \sin(\Theta/2);
	\\
	\ket{\psi_S} &= \ket{e_z}  
	\qq{when} 
	b_g \rightarrow -\sin(\Theta/2),\ b_e \rightarrow \cos(\Theta/2).
\end{equations}
Therefore, we get
\begin{equation}
	\ev{U^\dg \hn_A U}{g_z,\psi_M}= \ev{U^\dg \hn_B U}{e_z,\psi_M} = \frac{1}{2}\prtg{1 - \sin(\phi/2)\prtq{\cos[2](\Theta)+\Re{\ip{1_{\omega}}{1_{\omega+\omega_q}}}\sin[2](\Theta)}},
\end{equation}
which, inserted in Eq.~\eqref{APPeq:ErrorDecompositionSigmaZ} leads to Eq.~\eqref{eq:ErrorFormula} of the main text.
We can see that the error being independent of the qubit's initial state is a direct consequence of the equality of the two terms in Eq.~\eqref{APPeq:ErrorDecompositionSigmaZ}.

\clearpage

\section{Maximization of Ozawa's bound\label{APPSec:BoundMaximization}}

The error given in Eq.~\eqref{eq:ErrorFormula} of the main text is independent of the initial qubit's state. Therefore, it makes sense to compare it to Ozawa's bound [Eq.~\eqref{eq:OzawaBound} of the main text] maximized over all possible initial qubit states. 
This can be done as follows: first, we notice that $\comm{\sigma_z}{H_S} = i \hbar \omega_q \sin(\Theta) \sigma_y$ and inserting it into Ozawa's bound, which we report here:
\begin{equation}\label{APPeq:OzawaBound}
	\varepsilon^2 (\ket{\psi_S}) \geq \varepsilon^2_{\mathrm{B}} (\ket{\psi_S})
	=  \frac{1}{4} \frac{\abs{\ev{\comm{O_S}{L_S}}{\psi_S}}^2}{\Delta L_S^2 + \Delta L_M^2}.
\end{equation}
Recalling that $O_S=\sigma_z$ and upon defining $\sigma_\Theta\equiv \cos(\Theta)\sigma_z + \sin(\Theta)\sigma_x = L_S = H_S$, one gets
\begin{equation}
	\varepsilon_{\mathrm B}^2 \prt{\ket{\psi_S}} =
	\frac{\sin[2](\Theta) \ev{\sigma_y}^2}{1+4(\Delta \omega/\omega_q)^2 - \ev{\sigma_\Theta}^2}.
\end{equation}
In the above equation, we have used that $\Delta L_M = \Delta H_M = \hbar \Delta \omega$, with $\Delta \omega$ being the frequency dispersion of the flying particle.
For a fixed value of $\ev{\sigma_y}$, the goal is to maximize $\ev{\sigma_\Theta}^2$~\footnote{Splitting the maximization in two steps poses no problems because for such simple sets it holds the following: the maximum of the set of maxima of subsets covering the starting set is also the maximum of the entire set. For the same reason, the order of the two maximizations does not matter}. When this is done, we obtain that $\ev{\sigma_\Theta}^2 + \ev{\sigma_y}^2=1$, yielding
\begin{equation}\label{APPeq:OzawaBoundQubit}
	\varepsilon_{\mathrm B}^2=
	\max_{\ev{\sigma_y}}
	\frac{\sin[2](\Theta) \ev{\sigma_y}^2}{4(\Delta \omega/\omega_q)^2 +\ev{\sigma_y}^2}
	=
	\frac{\sin[2](\Theta)}{1 + 4 (\Delta \omega/\omega_q)^2},
\end{equation}
where the maximum is attained by choosing $\ev{\sigma_y}=\pm 1$, which implies $\ev{\sigma_\Theta} = 0$. The above formula shows that Ozawa's bound is now a simple function of the energy dispersion of the meter with respect to the energy scale of the measured system.

\clearpage
\section{Ratio between error and bounds for different wavepackets\label{APPSec:WavepacketShapesErrorBounds}}

In this section, we study the ratio $\varepsilon/\varepsilon_{\mathrm{B}}$ for different wavepacket shapes: a Gaussian wavepacket $\psi_\mathrm{G} (x)$, a square pulse wavepacket $\psi_\mathrm{Sq} (x)$, and an exponential decay wavepacket $\psi_\mathrm{Exp} (x)$. In order to make the notation less cumbersome we define the time variable $t = -x/v_0$. To simplify the formulas, we take Gaussian and square pulse wavepackets to be centered around $t=0$. Regarding the square and exponential wavepackets, their discontinuity causes problems in the calculation of $\Delta \omega^2$. For this reason we use instead smooth wavefunctions which approximate them. We write
\begin{equation}
	\psi_\mathrm{G} (t) = \frac{\exp[-\frac{t^2}{4 \sigma_t^2}]}{(2\pi)^{1/4}\sqrt{\sigma_t}},
	\quad
	\psi_\mathrm{Sq} (t) = \sqrt{\frac{\tanh(\frac{t+s}{\epsilon s}) - \tanh(\frac{t-s}{\epsilon s})}{4 s}},
	\quad
	\psi_\mathrm{Exp} (t) = \sqrt{\frac{2 \gamma}{\pi \epsilon}\sin(\frac{\pi \epsilon}{2}) \prtq{\frac{1+\tanh(\gamma t/\epsilon)}{2}}e^{-\gamma t}}
\end{equation}
where $2 s$ indicates, more or less, the temporal length of the square wavepacket while $\gamma$ is the decay rate of the exponential decay wavepacket. 
The square pulse and exponential decay wavepackets assume their idealized form in the limit $\epsilon \rightarrow 0^+$, but we will see that in this case their frequency variance diverges. Therefore, we will have to choose a finite value for the a-dimensional quantity $\epsilon$. As written, all wavepackets are normalized for any value of $\epsilon < 1$.

Calculating the time and frequency dispersions for the three wavepackets we get
\begin{equations}\label{APPeq:Dispersions}
	&\Delta t_\mathrm{G} = \sigma_t,&
	\qquad
	&\Delta t_\mathrm{Sq} = \frac{\sqrt{4+\pi^2\epsilon^2}}{2\sqrt{3}}s,&
	\qquad
	&\Delta t_\mathrm{Exp} = \frac{\pi \epsilon}{2 \gamma \sin(\pi \epsilon/2)},&
	\\
	&\Delta \omega_\mathrm{G} = \frac{1}{2\sigma_t},&
	\qquad
	&\Delta \omega_\mathrm{Sq} = \frac{\sqrt{\epsilon\sinh(4/\epsilon) -4}}{2\sqrt{2}\sinh(2/\epsilon)\epsilon s},&
	\qquad
	&\Delta \omega_\mathrm{Exp} = \sqrt{\frac{2-\epsilon}{8 \epsilon}}\gamma.&
\end{equations}
We can observe how the frequency dispersion of the square pulse and exponential decay wavepackets diverge for $\epsilon \rightarrow 0$ while their time dispersion does not.

Since we are interested in the effect that using different wavepackets shapes has on the ratio $\varepsilon/\varepsilon_\mathrm{B}$ we consider the error as always computed for the optimal value $\phi=\pi$ so that we get [cf. Eq.~\eqref{eq:ErrorFormulaOptimized} and Eq.~\eqref{eq:OzawaBoundQubit} of the main text]
\begin{equation}
	\label{APPeq:ErrorFormulaRatio}
	\frac{\varepsilon\vert_{\phi=\pi}}{\varepsilon_\mathrm{B}} = 
	\sqrt{2\prt{1+4\frac{\Delta \omega^2}{\omega_q^2}}(1-P)},
	\qq{where}
	P 
	=
	\intmp \dd{t} \abs{\psi_0 (t)}^2 \cos(\omega_q (t+\tau)),
\end{equation}
where $\tau$ represents our liberty of choosing the initial position of the wavepacket. We will always choose $\tau$ in order to maximize $P$.
Calculating $P$ for the three different cases gives
\begin{equations}
	P_\mathrm{G} &= 
	\max_\tau \prtg{\cos(\omega_q \tau )\exp(-\frac{1}{2}\omega_q^2 \sigma_t^2)}
	=\exp(-\frac{1}{2}\omega_q^2 \sigma_t^2),
	\\
	P_\mathrm{Sq} &= 
	\max_\tau \prtg{\frac{\epsilon}{4}\cos(\omega_q \tau) \prt{B_1+B_2+B_3+B_4}}
	= \frac{\epsilon}{4}\abs{B_1+B_2+B_3+B_4}
	\\
	P_\mathrm{Exp} &= 
	\frac{\sin(\pi \epsilon/2)}{4}\max_\tau \prtg{e^{-i \omega_q \tau} \prtq{\cot(\frac{\pi \epsilon(\gamma + i \omega_q)}{4\gamma})} + \tan(\frac{\pi \epsilon(\gamma + i \omega_q)}{4\gamma}) 
		+ 2 e^{2 i \omega_q \tau} \csc(\frac{\pi \epsilon(\gamma - i \omega_q)}{2\gamma})}.
\end{equations}
where
\begin{equations}
	B_1 &= e^{-i \omega_q s}e^{-(\pi/2)\epsilon \omega_q s} \int_0^{-e^{-2/\epsilon}} \frac{k^{+(1/2)i \epsilon \omega_q s}}{1-k}\dd{k},
	\qquad
	B_2 &= e^{+i \omega_q s}e^{+(\pi/2)\epsilon \omega_q s} \int_0^{-e^{-2/\epsilon}} \frac{k^{-(1/2)i \epsilon \omega_q s}}{1-k}\dd{k},\\
	B_3 &= e^{+i \omega_q s}e^{-(\pi/2)\epsilon \omega_q s} \int_0^{-e^{2/\epsilon}} \frac{k^{+(1/2)i \epsilon \omega_q s}}{1-k}\dd{k},
	\qquad
	B_4 &= e^{-i \omega_q s}e^{+(\pi/2)\epsilon \omega_q s} \int_0^{-e^{2/\epsilon}} \frac{k^{-(1/2)i \epsilon \omega_q s}}{1-k}\dd{k}.
\end{equations}
In the case of the exponential decay wavepacket, the maximization over $\tau$ does not seem possible to be made analytically, so we will do it numerically.

\begin{figure}
	\centering
	\includegraphics[width=0.7\textwidth]{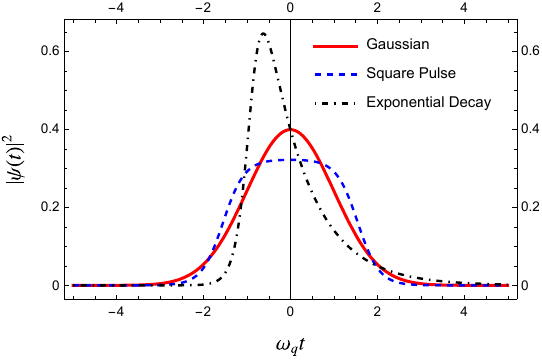}
	\caption{Shape of $\abs{\psi_\mathrm{G} (t)}^2$,$\abs{\psi_\mathrm{Sq} (t)}^2$, and $\abs{\psi_\mathrm{Exp} (t)}^2$ for $\epsilon=1/\pi$ and $\omega_q \Delta t = 1$. All curves are plotted so that they are centered around zero. We can see how already for this value of $\epsilon$ we get shapes resembling an ideal square pulse and an ideal exponential decay.}
	\label{APPfig:WavepacketsPlots}
\end{figure}

Now we have all the ingredients to compute the error ratio as a function of the single a-dimensional parameter $\omega_q \Delta t$. In every formula we substitute $s$ and $\gamma$ by inverting the correspondent time-dispersion equations in~\eqref{APPeq:Dispersions}. Then we also write the frequency dispersions as functions of $\Delta t$. Regarding $\epsilon$, we choose it to be equal to $1/\pi$. As one can see in fig.~\ref{APPfig:WavepacketsPlots}, this value already allows for a good approximation of the ideal shapes while maintaining a finite $\Delta \omega$.
The resulting plot is the one reported in the main text.

\end{document}